\documentclass[manuscript]{aastex}
\usepackage{rotating}
\usepackage{tikz}
\usepackage{amsmath}

\usepackage{graphicx}

\usepackage{pdflscape}
\usepackage[normalem]{ulem} % either use this (simple) or
\usepackage{soul} % use this (many fancier options)
\usepackage{aas_macros}

\slugcomment{\textbf{Revised 2026 June 18} }

\shorttitle{240P}
\shortauthors{Jewitt}

\begin{document}

\title{Investigation of Split Comet 240P/NEAT}
%% Use \author, \affil, and the \and command to format
%% author and affiliation information.
%% Note that \email has replaced the old \authoremail command
%% from AASTeX v4.0. You can use \email to mark an email address
%% anywhere in the paper, not just in the front matter.
%% As in the title, use \\ to force line breaks.

\author{
David Jewitt$^{1}$, 
Jane Luu$^{2}$ and Yoonyoung Kim$^{1}$
} 
\affil{$^1$Department of Earth, Planetary and Space Sciences, UCLA, 595 Charles Young Drive, Los Angeles, CA 90095\\}
\affil{$^2$ Centre for Planetary Habitability (PHAB), Department of Geosciences, University of Oslo, NO-0315 Oslo, Norway}

\email{djewitt@gmail.com}

\begin{abstract}
We present  time-series observations of the split comet 240P/NEAT near perihelion, obtained using the Nordic Optical Telescope.   The brighter component, 240P-A, has an estimated radius in the range 400 m to 600 m, and loses dust at the peak rate 130 kg s$^{-1}$.  The ejected dust has characteristic size $\sim$50 $\mu$m,  is expelled sunward at $\sim$25 m s$^{-1}$, with a total ejected mass in the period of observation $\sim1.5\times10^9$ kg.  Mass loss from the fainter component, 240P-B, peaks at $\sim$35 kg s$^{-1}$ and the total ejected mass was 2.3$\times10^8$ kg.  The radius of 240P-B is uncertain, with a best estimate $\sim$300 m and an absolute lower limit 50 m.   240P-A and 240P-B are currently separating at about 1 m s$^{-1}$, a speed that is likely accelerating as a result of differential outgassing forces, and have a separation age of $>$3 years. The splitting of 240P is incompatible with the action of tides, impact, and internal pressure build up. 240P fits a developing picture, in which small comets are destroyed by rotational instabilities triggered by outgassing torques, an explanation that can be tested in 240P by future observations.

\end{abstract}

%\keywords{comets: general---comets: individual (C/2021 O3) }

\section{INTRODUCTION}
\label{intro}

240P/NEAT (hereafter 240P) is a typical Jupiter family comet, with  orbital semimajor axis $a$ = 3.862 au, eccentricity $e$ = 0.450 and inclination $i$ = 23.5\degr~(Tisserand parameter, $T_J$ = 2.758).  Discovered only in 2002 \citep{Law02}, 240P is most likely a recent arrival from the Kuiper belt.   Perihelion last occurred on UT 2025 December 19.9 (Day of Year DOY=354) at distance $q$ = 2.12 au, while the next aphelion will be on 2029 September 30 at $Q$ = 5.60 au.  The orbit brings 240P in and out of the water ice sublimation zone, with blackbody subsolar temperature varying on the 7.6 year orbital timescale from 395 K at perihelion to 166 K at aphelion.   240P experiences frequent, strong interactions with Jupiter (semi-major axis 5.20 au), the most recent  being an approach to within 0.25 au on 2007 July 9.  The current activity of 240P, including several long-lived photometric outbursts in 2018/2019, might be related to a reduction in the perihelion distance (from 2.5 au to 2.1 au) brought about by the  encounter in 2007 \citep{Kel19}.  A fainter co-moving object, 240P-B\footnote{In the following, we use the names 240P-A and 240P-B as allocated by the Minor Planet Center, without prejudice to which of these might be the precursor (more massive) body.}, was first reported on UT 2025 June 4 \citep{Jae25}, at which time the comet was inbound at 2.6 au.    

While numerous split comets have been recorded over the last century (see reviews by \cite{Sek82} and \cite{Boe04}), detailed physical studies of such objects only rarely find their way into the refereed literature.  This is unfortunate because growing evidence suggests that splitting and disintegration, not sublimation losses, are the primary mechanisms of cometary destruction, at least at small nucleus sizes (\cite{Jew97}, \cite{Sam07}, \cite{Jew21}). 
Our purpose here is first, to put  new observations of split comet 240P on record, by examining the photometric and morphological evolution of the two components and second, to establish a likely cause of the nucleus splitting.

\section{OBSERVATIONS}
Observations were taken using the Alhambra Faint Object Spectrograph and Camera (ALFOSC) at the f/11 Cassegrain focus of the 2.56 m diameter Nordic Optical Telescope. In imaging mode, the ALFOSC offers a $\sim$6\arcmin~unvignetted field of view with 0.214\arcsec~pixels.  For the present purposes, we employed a Bessell R filter \citep{Bes90}, which has a 1480\AA~passband width (Full Width at Half Maximum) and a central wavelength 6100\AA.  Bias and flat field exposures were obtained each night, the latter using a diffusely illuminated screen inside the telescope dome.  Calibration of the data was obtained using Landolt stars \citep{Lan92} and checked against calibrated field stars in the comet fields.  

\begin{figure}
\epsscale{0.99}

%\plotone{RDa.pdf}
\plotone{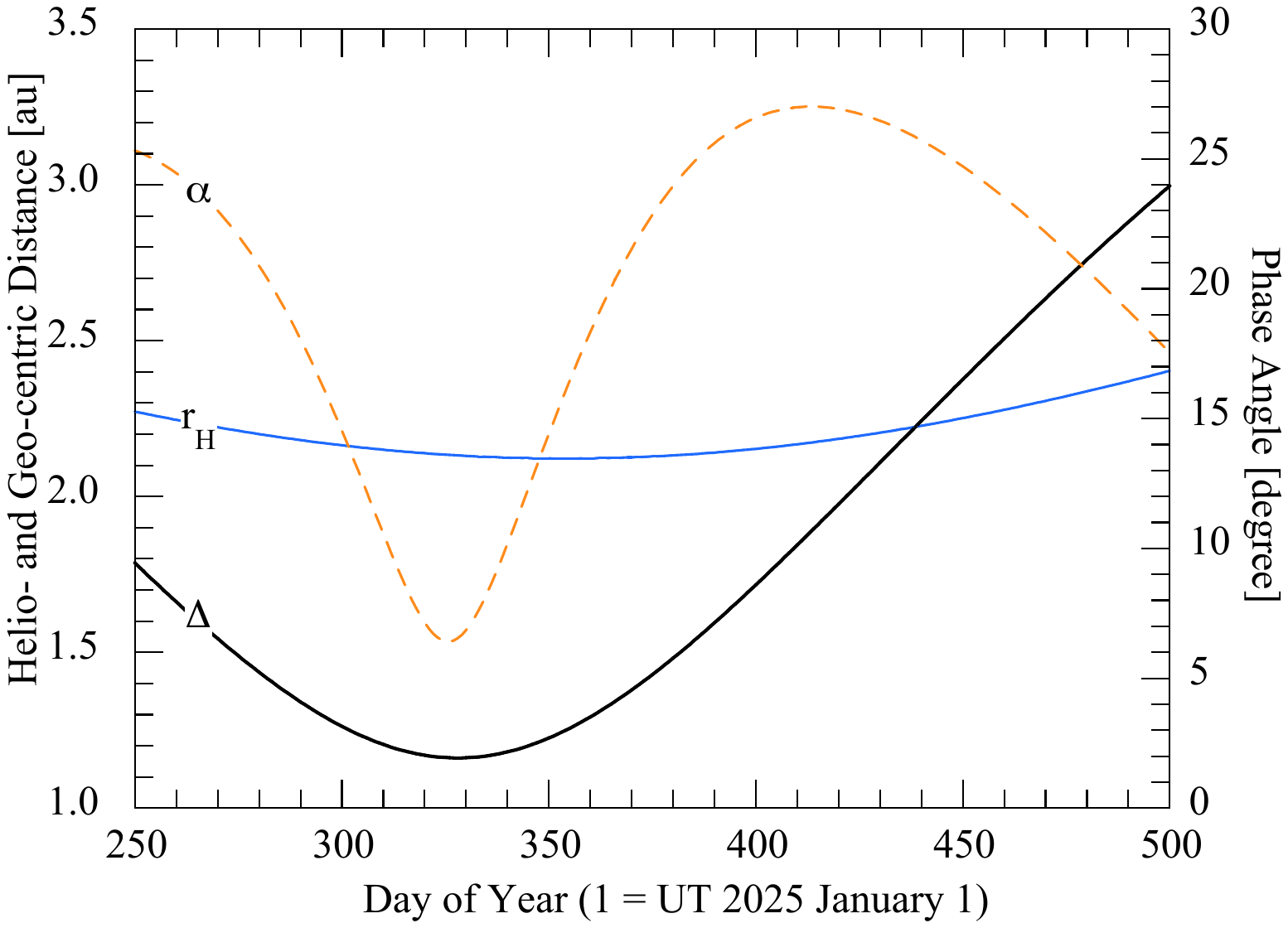}

\caption{Geometry of observation as a function of observation date.  Heliocentric ($r_H$) and geocentric ($\Delta$) distances refer to the left hand axis while phase angle ($\alpha$) is plotted on the right.  \label{RDa}}
\end{figure}

A journal of observations is given in Table \ref{240P_geometry} and the variation of the observing geometry is shown in Figure \ref{RDa}.  Our observations record the comet from 2 months before perihelion to 4 months after, during which time the heliocentric distance changed only slightly (2.12 au  to 2.29 au), corresponding to only a $\sim$16\% change in the insolation. 

\subsection{Morphology}
\begin{sidewaysfigure}
%\rotate
\epsscale{0.99}
%\plotone{composite.pdf}
\plotone{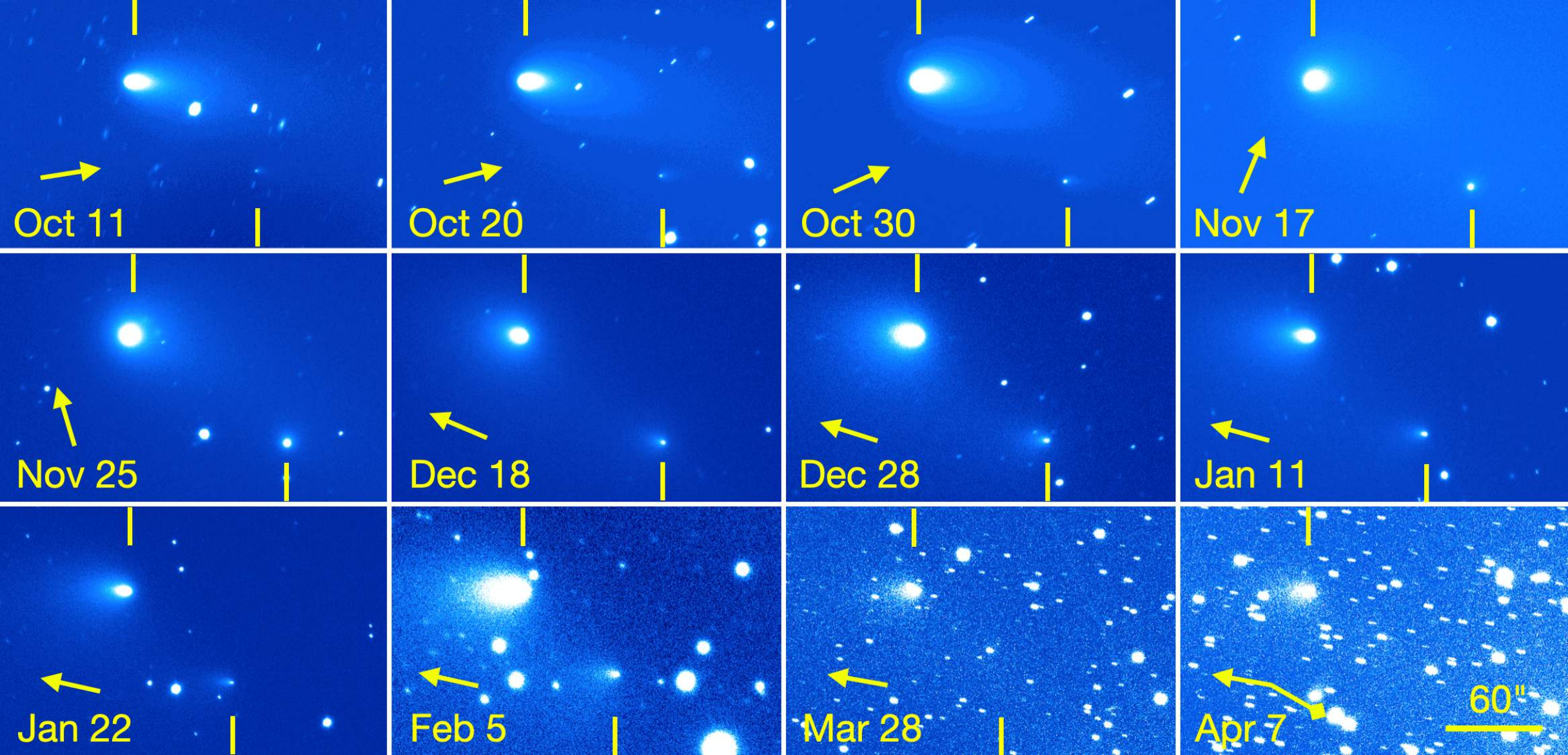}

\caption{Composite of images showing the development of 240P from 2025 October 11 to 2026 April 7 (c.f., Table \ref{240P_geometry}).  240P-A and 240P-B are marked by short vertical lines above and below the objects, respectively. Arrows show the antisolar direction (-S) at each epoch.  The negative projected velocity vector (-V) barely changes from panel to panel and is shown only in the April 7 panel for clarity.  North is to the top, East to the left, and each panel shows a region 240\arcsec~wide.  \label{composite}}
%\end{figure}
\end{sidewaysfigure}

Figure \ref{composite} shows the evolution of the morphology of 240P from each date of observation listed in Table \ref{240P_geometry} (except UT 2025 December 2, omitted because the seeing then was unusually poor).
The teardrop shaped brighter component of the comet, 240P-A, is obvious and marked by a short vertical line at the top of each panel.  The fainter component, 240P-B, is indicated by a short vertical line at the bottom.  Arrows in the figure show the antisolar direction, which swings progressively from the west to the east because of the changing viewing geometry.  

\begin{figure}
\epsscale{0.99}

%\plotone{angle_plot.pdf}
\plotone{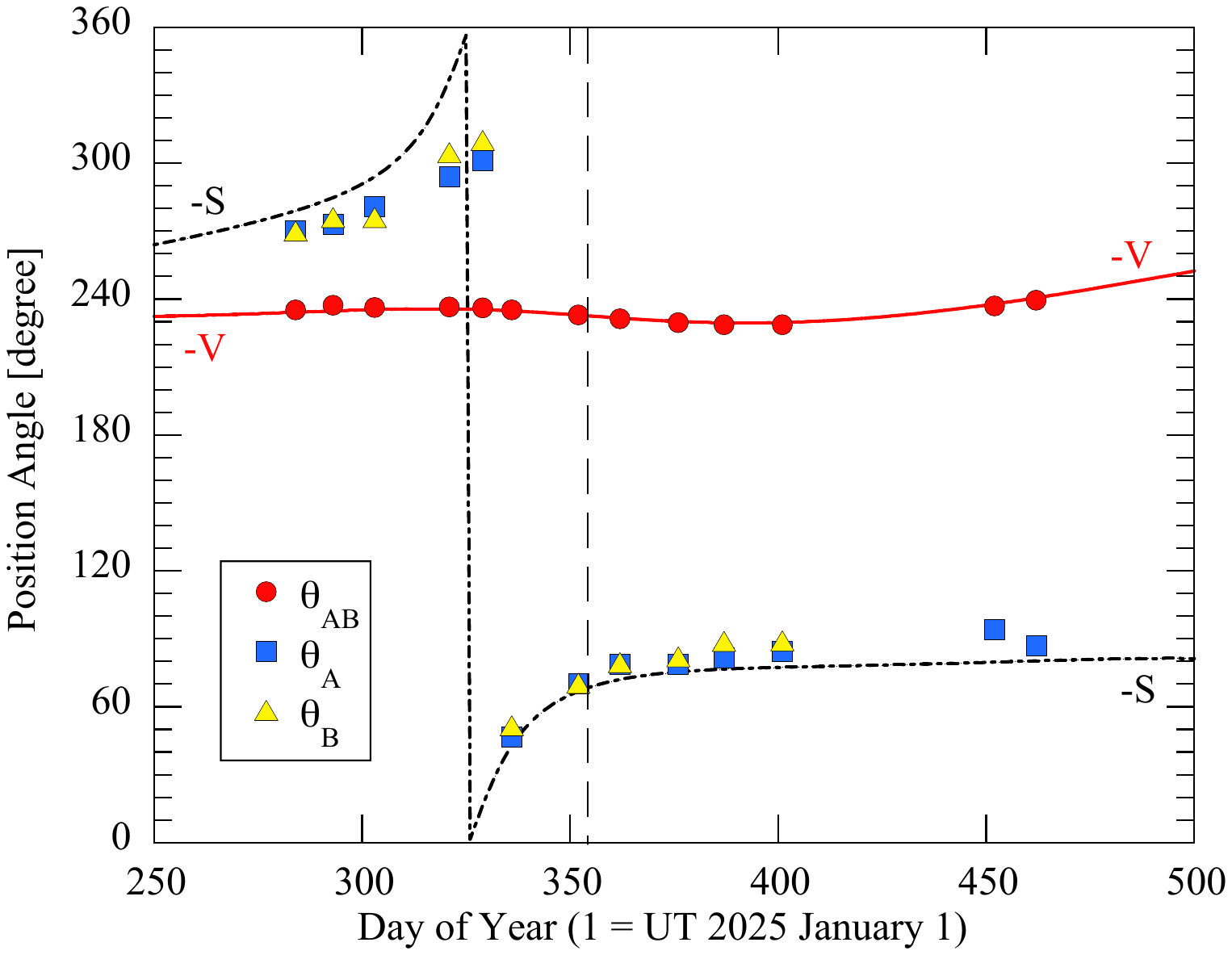}

\caption{Measured position angles of the line connecting 240P-A to 240P-B line (red circles labelled $\theta_{AB}$) and of the tail directions on 240P-A and 240P-B (blue squares and yellow triangles, labelled $\theta_A$ and $\theta_B$, respectively).  The solid red line shows the negative projected velocity of the comet while the dot-dash black line shows the projected anti-solar direction.  A vertical dashed black line marks the date of perihelion. \label{angle_plot}}
\end{figure}

Figure \ref{angle_plot} shows $\theta_{AB}$ (red filled circles), the position angle of a line drawn from 240P-A  to 240P-B,  compared with the position angle of the projected negative velocity vector (solid red line marked ``-V'') computed from JPL Horizons.  The  correspondence between $\theta_{AB}$ and the model indicates that 240P-A and 240P-B share approximately the same orbit plane.   Also shown in the figure are the position angles of the tails on 240P-A (blue filled squares marked $\theta_A$) and 240P-B (yellow filled circles marked $\theta_B$), compared with the projected anti-solar direction (shown as a dashed black line marked ``-S'').  Since the tails are diffuse and curved (i.e., $\theta_A$ and $\theta_B$ vary with distance from their respective nuclei) we have systematically measured the position angle at a fixed distance (20 pixels) from the nucleus on each date.  The figure shows a) that the tail position angles vary together while remaining closely parallel and b) that they follow the angular variations in $-S$, albeit with an offset from the anti-solar direction.  These properties are expected of particles small enough to respond to the changing direction of solar radiation pressure. 

\subsection{Photometry}
\begin{figure}
\epsscale{0.7}

%\plotone{apparent.pdf}
%\plotone{absolute.pdf}
\plotone{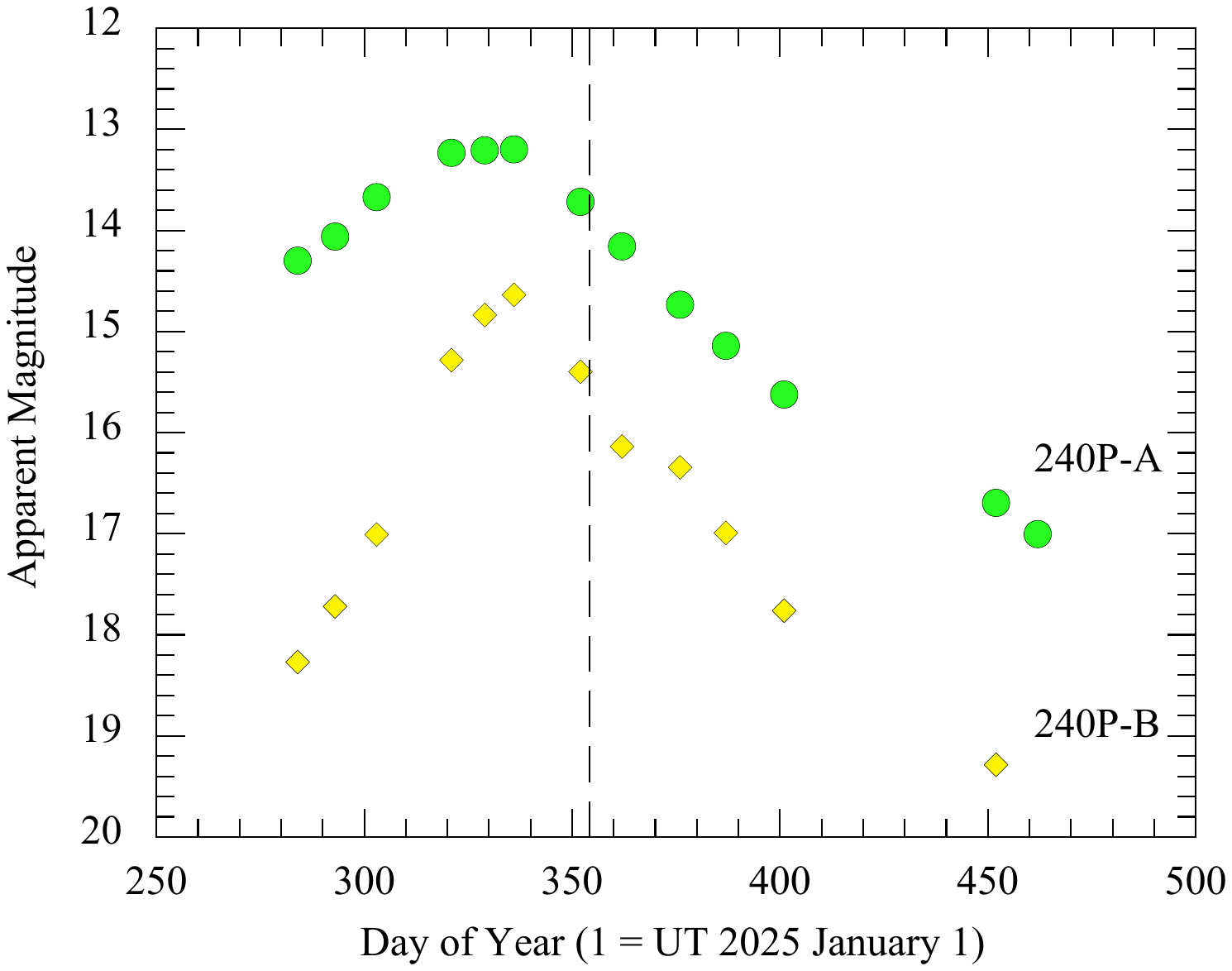}
\plotone{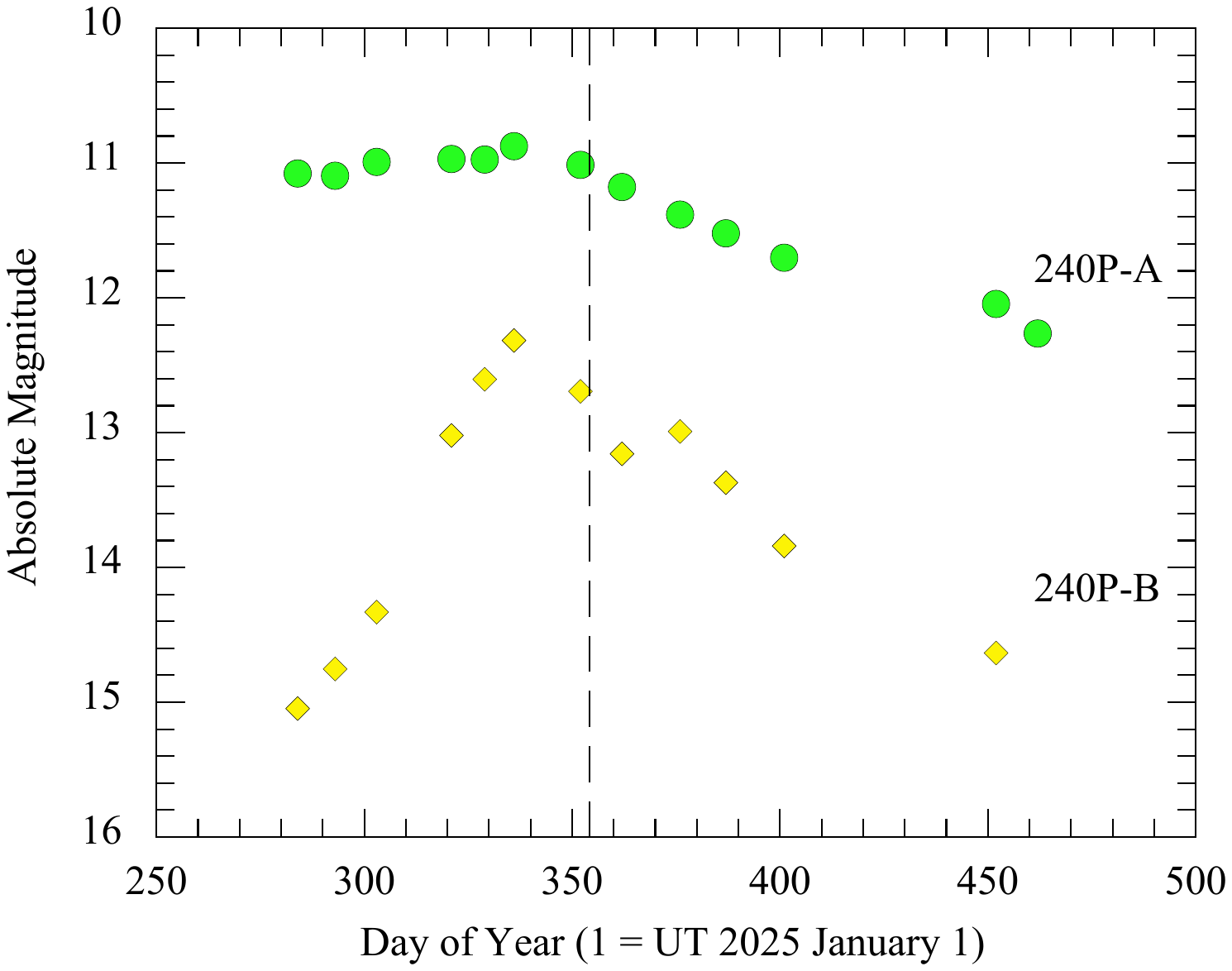}
\caption{(Upper:) Apparent R band magnitudes of 240P-A (green circles) and 240P-B (yellow diamonds) measured within an aperture of fixed linear radius 10$^4$ km.  Error bars are smaller than the symbols used to show the data.  (Lower:) Absolute magnitude as a function of observation date. In both panels, the vertical dashed line marks the date of perihelion. \label{apparent}}
\end{figure}

We measured the magnitudes of 240P-A and 240P-B using apertures having  fixed linear radii of 10$^4$ km when projected to the distance of the comet.  Use of a fixed linear (as opposed to angular) radius aperture samples the material within a fixed volume around each nucleus, and has the advantage of removing complications in the interpretation of the photometry caused by changing geocentric distance and the complex coma morphology.  Sky subtraction was obtained using the median signal computed within a concentric annulus having projected inner and outer radii 100 and 300 pixels (21.4\arcsec~ and 64.2\arcsec), respectively, with additional spot checks at other star-free locations in the surrounding field.  The selection of the sky annulus was a compromise; smaller sky annuli suffered more from dust tail contamination while larger apertures suffered increasingly from small deviations in the flatness of the reduced images.  The resulting apparent magnitudes are shown in the upper panel of Figure \ref{apparent}. 

\begin{figure}
\epsscale{0.95}

%\plotone{difference.pdf}
\plotone{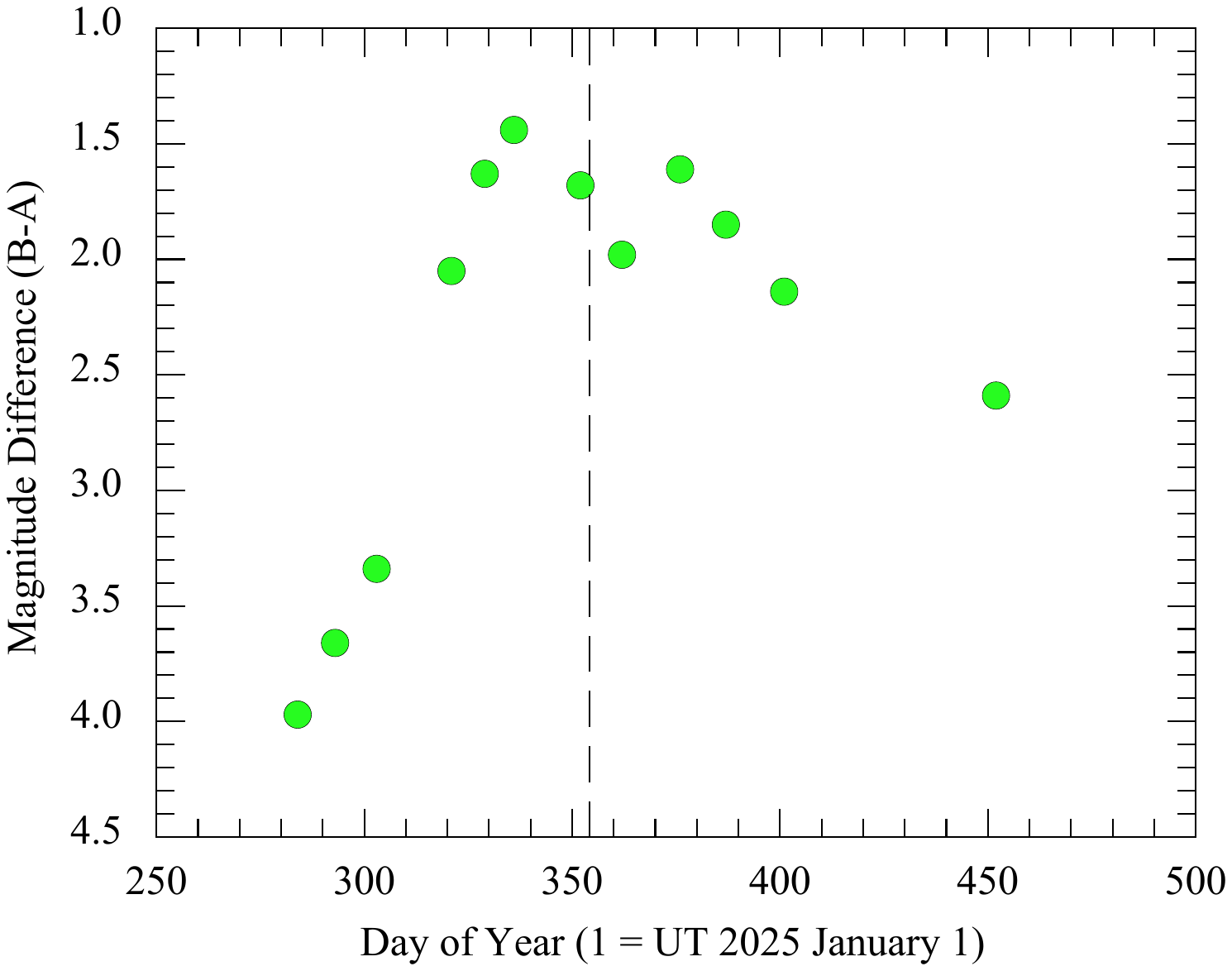}

\caption{Difference of the 10$^4$ km aperture magnitudes, B-A, vs.~date of observation.  The vertical dashed line marks the date of perihelion.\label{difference}}
\end{figure}

The apparent magnitudes reflect both temporal changes in the quantity of scattering material around the nucleus, and in the observing geometry.  We corrected for the latter using the inverse square law, written 

\begin{equation}
H = m_R - 5\log_{10}(r_h \Delta) - \beta_{\alpha} \alpha
\end{equation}

\noindent in which $r_H$ and $\Delta$ are the heliocentric and geocentric distances, $\alpha$ is the phase angle and $m_R$ is the apparent red magnitude.  Quantity $\beta_{\alpha}$ is the phase function, which is unmeasured in 240P and which we assume to be $\beta_{\alpha}$ = 0.04 magnitude degree$^{-1}$ across the 6\degr~$\le \alpha \le$ 27\degr~range of phase angles of the present data (c.f., Table \ref{240P_geometry}).  
The resulting absolute magnitudes are shown in the lower panel of Figure \ref{apparent}, where it is apparent that both components varied with time, but that 240P-B varied more.  The magnitude difference (B-A) is plotted as a function of date in Figure \ref{difference}.  Relative to 240P-A, 240P-B brightened by $\sim$2.5 magnitudes (factor of 10) in two months before reaching a peak 15 days before perihelion, then faded by $\sim$1 magnitude (factor 2.5) for three months after perihelion. The different responses to a common insolation presumably reflect different distributions of ice on the two objects. For example, the two bodies might have different pole directions and so experience different seasonal illumination effects.  Ice might also be distributed differently with depth  within the two bodies and stochastic processes like cliff collapse, which are known to occur on comets \citep{Dav24}, could alter the surface ice distribution.  

The effective scattering cross-section, $C$, is related to absolute magnitude by

\begin{equation}
C = \frac{1.5\times10^6}{p_R} 10^{-0.4 H}
\label{crosssection}
\end{equation}

\noindent in which $p_R$ is the red geometric albedo of the material, assumed to be $p_R$ = 0.04.  The apparent and absolute magnitudes and the related cross-sections are summarized in Table \ref{240P_data}.

\subsection{Particle Size}
\label{finson}
The motion of a dust particle under the simultaneous action of solar gravitational acceleration and radiation pressure is defined by the dimensionless parameter, $\beta$, equal to the ratio of the accelerations induced by solar photons and by gravity \citep{Fin68}.  The locus of positions of particles having a range of $\beta$ values but released from the nucleus at a single time (e.g., as in a cometary outburst) is called a synchrone.  The locus of positions of particles having one value of $\beta$ but released over a range of times is called a syndyne.  For spherical dielectric particles larger than the wavelength of light, the $\beta$ parameter is approximately related to the particle radius by $\beta \sim 1/a_{\mu}$, where $a_{\mu}$ is the radius expressed in microns (e.g., \cite{Boh83}).  Cometary particles are irregularly shaped aggregates, not spheres, but Finson-Probstein models of spherical grain dynamics nevertheless provide  a useful and widely employed representation of dust motions in real comets.

\begin{figure}
\epsscale{0.99}
%\plotone{pair.pdf}
%\plotone{synsyn.pdf}
\plotone{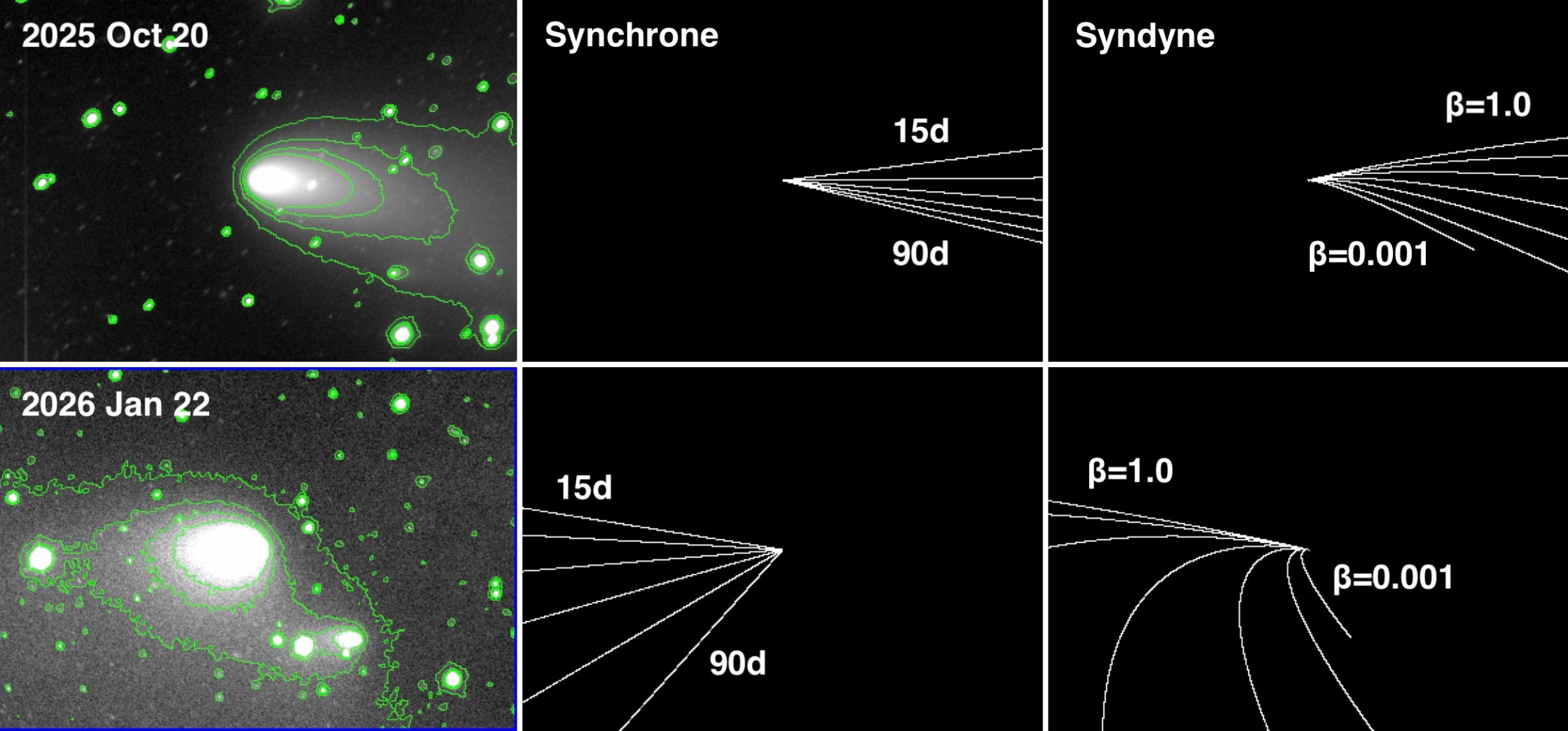}

\caption{NOT images from (upper) UT 2025 October 20 and (lower) UT 2026 January 22.  Synchrone and syndyne trajectories are shown in the middle and right-hand panels for each epoch.  \label{synsyn}}
\end{figure}

Figure \ref{synsyn} shows  synchrone and syndyne trajectories for images taken before perihelion on UT 2025 October 20 and after perihelion on UT 2026 January 22.  The synchrones describe particles ejected 15, 30, 45, 60, 75, 90 days before each observation, while the syndynes show particles having $\beta$ =  0.001, 0.003, 0.01, 0.03, 0.1, 0.3 and 1, corresponding to radii from microns to millimeters.  Inspection shows that the curved tail in the images is better matched by the syndynes than by the (linear) synchrones.  Of the former, $\beta$ = 0.001 is clearly too small and $\beta$ = 1 too large, with the best match being $\beta$ = 0.01 to 0.03.  Corresponding particle radii lie in the range 30 $\mu$m to 100 $\mu$m.  These estimates are approximate, not least because the Finson-Probstein model assumes that dust is released from the nucleus isotropically and at zero speed; neither assumption is likely to be true in practice.  It is also possible that ejected dust particles may be subject to erosion, through the loss of embedded volatiles or through electrostatic or rotational bursting, changing their effective $\beta$ as they recede from the nucleus.  With these caveats in mind, we take 0.01 $< \beta <$ 0.03 as our best guess as to the particle size, with the middle value, $\beta$ = 0.02, indicating an effective mean particle radius $\bar{a}$ = 50 $\mu$m.

\subsection{Particle Speed}
\label{speed}
Grains ejected towards the Sun are turned around by radiation pressure, giving the sunward coma an umbrella or nose-like shape.  The sunward extent of the coma, $X_R$, is related to the speed of ejection of the grains, $V_d$,  by 

\begin{equation}
\frac{V_d^2}{\beta} = 2 \left(\frac{g(1)}{r_H^2}\right) X_R
\label{nose}
\end{equation}

\noindent where $g(1) = 6\times10^{-3}$ m s$^{-2}$ is the solar gravitational acceleration at $r_H$ = 1 au, $r_H$ is the heliocentric distance measured in au and $\beta$ is the radiation pressure efficiency \citep{Jew87}.  If the apex of the dust motion lies on the Sun-comet line, then the apparent length of the ``nose'' in the plane of the sky is $X_R^{'} = X_R \sin(\alpha)$, where $\alpha$ is the phase angle. 

From each image we measured the surface brightness profile along the Sun direction, with an extraction box of perpendicular width  20 pixels (4.3\arcsec).  We measured the distance between the peak surface brightness and the location where the surface brightness fell to 10\% of the peak in order to define  $X_R$.  Images for which phase angle $\alpha <$ 12\degr~were rejected to avoid excessive $1/\sin{\alpha}$ projection correction factors at smaller angles. For  240P-A, the resulting mean value is  $X_R^{'} = (4.4\pm0.5)\times10^6$ m (median 4.2$\times10^6$ m) for the plane-of-sky nose length and  $X_R = (12.9\pm1.3)\times10^6$ m (median 13.0$\times10^6$ m) for the deprojected value.  The fainter object 240P-B has much larger uncertainty but the same median nose length.  Substitution into Equation \ref{nose} gives the mean value of $V_d^2/\beta = (3.3\pm0.3)\times10^4$ m$^2$ s$^{-2}$.  Substituting $\beta$ = 0.02 (as found from the Finson-Probstein analysis in section \ref{finson}) we find  $V_d$ = 25 m s$^{-1}$, as a working value appropriate for $\sim$50 $\mu$m radius dust grains.

\subsection{Production Rates}
\label{production}
\begin{figure}
\epsscale{1.0}

%\plotone{dmbdt_vs_doy.pdf}
\plotone{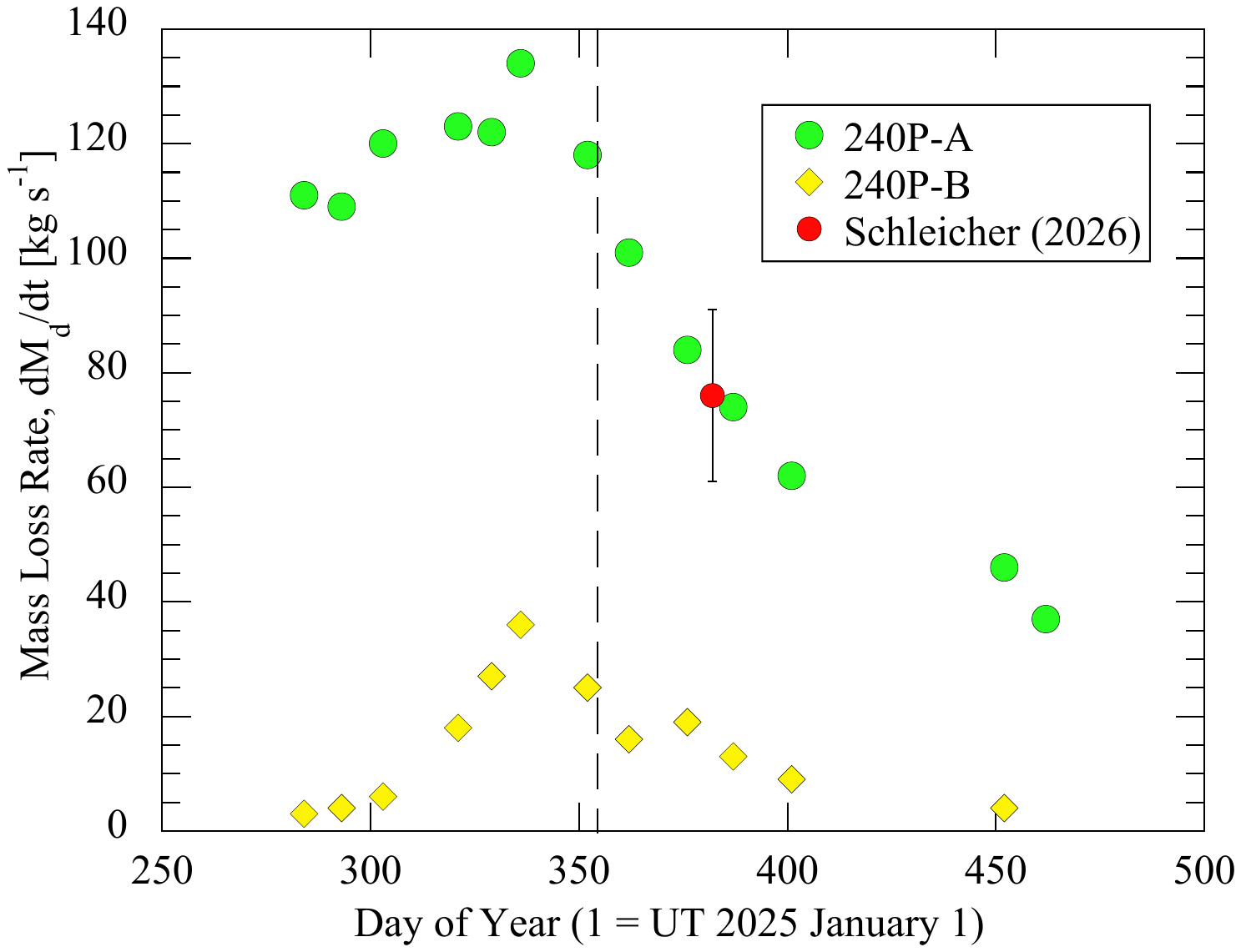}

\caption{Dust mass loss rates from 240P-A (green circles) and 240P-B (yellow diamonds) (c.f., Table \ref{240P_data}).  The red circle shows the water production rate by Schleicher (2026).  The dashed vertical line marks the date of perihelion.  \label{dmbdt_vs_doy}}
\end{figure}

The photometry and the particle size and speed estimates derived from the morphology together give an estimate of the production rate in dust.  The mass of an optically thin assemblage of spheres is related to the sum of their cross-sections, $C$, by $M_d = 4\rho \bar{a} C/3$, where $\rho$ is the particle density and $\bar{a}$ the mean radius.  The time taken for particles to cross a photometry aperture of projected radius $L$ is $t = L/V_d$.  The ratio gives the mass production rate

\begin{equation}
\frac{dM_d}{dt} = \frac{4\rho \bar{a} C V_d}{3 L}
\label{dmbdt}
\end{equation}

\noindent measured in kg s$^{-1}$.  Values of $dM_d/dt$ are listed for both 240P-A and 240P-B in Table \ref{240P_data} and plotted in Figure \ref{dmbdt_vs_doy}, all assuming $\bar{a}$ = 50 $\mu$m, $\rho$ = 500 kg m$^{-3}$, $V_d$ = 25 m s$^{-1}$ and $L$ = 10$^7$ m.  Uncertainties are considerable, including a factor of 2 on $\bar{a}$ and the fact that $\rho$ is unmeasured (and could easily be different by another factor 2).  The mass loss rate estimates should therefore be taken as highly uncertain.  On the other hand, David Schleicher (private communication, January 20, 2026) measured the water production rate from 240P on UT 2026 January 17/18 (DOY = 382/383; $r_H$ = 2.13 au) of $Q_{H_2O} = 2.5\times10^{27}$ s$^{-1}$, corresponding to a water mass production rate $\dot{M}$ = 76 kg s$^{-1}$.  The agreement with the (74 kg s$^{-1}$ to 84 kg s$^{-1}$; c.f., Table \ref{240P_data})  dust production rate at this time is, if anything, better than expected given the uncertainties.  

The integrated  mass of dust lost from 240P-A between the first (UT 2025 October 11) and last (UT 2026 April 7) NOT observations  is $\Delta M_d = 1.5\times10^9$ kg. This is a lower limit to the total mass lost per orbit because we based $\Delta M_d$ on observations from only a limited window around perihelion (c.f., Table \ref{240P_geometry}) and the comet was clearly active before our first, and after our last, observations.   Furthermore, as is true of any mass estimate based on scattered sunlight, it is always possible that mass could be hidden in a small number of large bodies that are individually too small to detect.

Measurements of 240P-B parallel those of 240P-A but are hindered by the poorer signal-to-noise ratios of the fainter component.  The curved morphology of the tail is the same in both objects, as is the median nose length for both objects, $X_R$ = 13$\times10^6$ m.  For these reasons, we assume that the effective particle size and sunward ejection speed are the same in 240P-B as in 240P-A, and we again used Equation \ref{dmbdt} to estimate the integrated mass loss from 240P-B, finding $\Delta M_d = 2.3\times10^8$ kg, or about 15\% of the loss from 240P-A.

\section{DISCUSSION}

%Rsun = -27.09 (from V = -26.74 and V-R = 0.35 (Holmberg et al 2005)

\subsection{Size of the Nucleus}
\label{size}

We combine the water mass production rate with a measurement of the non-gravitational acceleration to obtain an estimate of the size of the nucleus of 240P-A.  The JPL Horizons non-gravitational acceleration parameters of 240P-A are $A1= (2.96\pm0.06)\times10^{-8}$ au day$^{-2}$,   $A2= (2.51\pm0.05)\times10^{-8}$ au day$^{-2}$, where 1 au day$^{-2}$ = 20.09 m s$^{-2}$.  A3 is assumed to be equal to zero.  (Parameters $A1 = (3.63\pm0.09)\times10^{-8}$ and $A2 = (2.94\pm0.10)\times10^{-8}$ au day$^{-2}$ are independently reported by S. Nakano\footnote{\url{https://tinyurl.com/NEAT-A}}.  The fact that these independent estimates of A1 and A2 differ by much more than the reported uncertainties is a measure of the difficulty of the measurement and indicates that the errors are under-estimated). Combining the Horizons A1 and A2 in quadrature gives non-gravitational acceleration at 1 au, $\zeta(1) = (7.80\pm0.03)\times10^{-7}$ m s$^{-2}$.  

Assuming that $\zeta$ is recoil acceleration from anisotropic outgassing we can estimate the radius of an equal mass sphere using \citep{Jew22}

\begin{equation}
r_n = \left(\frac{3 k_R V_{th} \dot{M(1)}}{4\pi \rho_n \zeta(1)}\right)^{1/3}.
\label{r_n}
\end{equation}

\noindent In Equation \ref{r_n}, $k_R$ is a dimensionless constant that accounts for the angular distribution of the mass loss ($k_R$ = 0 for isotropic mass loss, $k_R$ = 1 for perfectly collimated mass loss), $V_{th}$ is the speed with which the mass is lost, $\dot{M(1)}$ is the mass loss rate in water molecules at $r_H$ = 1 au, and $\rho_n$ is the nucleus density.

Scaling the water production rate, $\dot{M}$ = 76 kg s$^{-1}$ at 2.13 au to 1 au assuming equilibrium sublimation from the sun-facing hemisphere, we find a water production rate $\dot{M}(1)$ = 540 kg s$^{-1}$.   (This is within a factor of two of the  rate, $\dot{M}(1)$ = 380 kg s$^{-1}$, obtained using the size vs.~reduced magnitude relation of \cite{Jor92} with the photometry from Table \ref{240P_data}). The thermal speed of water molecules at $r_H \sim$2 au is $V_{th} = $ 500 m s$^{-1}$.  We also take $k_R$ = 1/2 \citep{Jew20} and $\rho_n$ = 500 kg m$^{-3}$ \citep{Gro19}, to find the radius of 240P-A as $r_n$ = 430 m.  It is difficult to assign a formal uncertainty to $r_n$ since $k_R$, $V_{th}$ and $\rho_n$ are unmeasured in 240P and do not themselves have formal uncertainties. We simply note that a factor of two error in $k_R V_{th}/\rho_n$ would give a $(2^{1/3} - 1) \sim$ 25\% error in $r_n$. Moreover, some fraction of $\dot{M}$ could be released by the sublimation of icy grains in the coma, in which case $r_n$ = 430 m would  represent an upper limit to the nucleus radius. 

A second estimate of the nucleus radius is obtained from the specific rate of sublimation  from the sunward hemisphere of a perfectly absorbing water ice nucleus at 2.13 au, namely, $f_s(1) = 3.1\times10^{-5}$ kg m$^{-2}$ s$^{-1}$.  At this rate, to generate water at rate $\dot{M}(2.13)$ = 76 kg s$^{-1}$ would require sublimation from a nucleus of radius 

\begin{equation}
r_n = \left(\frac{\dot{M}(r_H)}{2\pi f_s(r_H)}\right)^{1/2}
\label{r_n2}
\end{equation}

\noindent Substitution gives $r_n$ = 620 m, which is broadly consistent with the radius estimated from the non-gravitational acceleration. Equation \ref{r_n2} would set an upper limit to the radius of 240P-A if some fraction of the water production came from sublimating ice grains in the coma, but a lower limit  if the surface of the nucleus is only partially active.  The geometric cross-section of a 400 m to 600 m radius nucleus, $C \sim$ 0.5 to 1.1 km$^2$, is three orders of magnitude smaller than the photometrically measured cross-sections in Table \ref{240P_data}, confirming the dominance of dust. We take a middle value between the estimates from Equation \ref{r_n} and \ref{r_n2}, $r_n = 500\pm$100 m, for the radius of the nucleus of 240P-A.

The effective radius of 240P-B is much less certain.  The main observational clue is that the dust production rate from 240-B is always smaller than that from 240-A, but the difference is a function of time (Figure \ref{difference}) and strictly provides, at best, a measure of the ratio of active areas, not the ratio of component radii.  The minimum difference, $\Delta m \sim$ 1.5 magnitudes, would correspond to effective areas in the ratio 1:4 and suggests radius $r_n \sim$ 310 m, but 240P-B could be larger if the active fraction of the surface is substantially less than unity.  240P-B could even be larger than 240P-A, given the limited available information.  An absolute minimum to the radius of 240P-B is set by the integrated mass loss, $\Delta M_d = 2.3\times10^8$ kg, which is equal to the mass of a sphere with density $\rho_n$ = 500 kg m$^{-3}$ and radius $r_n$ = 48 m.

%PICK A DEFINITE RADIUS

\subsection{Separation of the Components}
\begin{figure}
\epsscale{0.8}

%\plotone{offset_plot.pdf}
\plotone{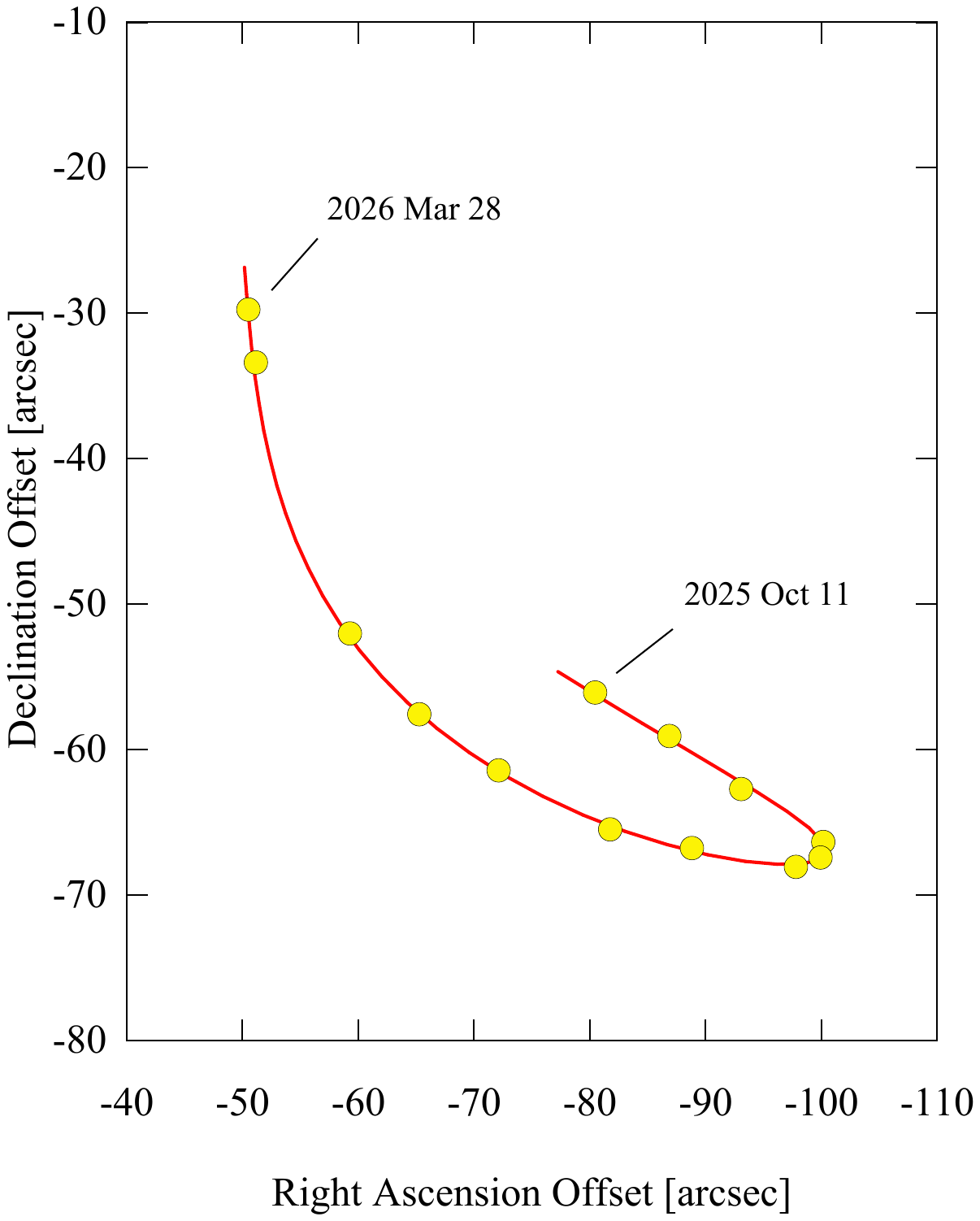}

\caption{Measured sky-plane offset of 240P-B from 240P-A (yellow filled circles) as a function of time, compared with the best-fit FindOrb solution (red line).  240P-B is South and West of 240P-A. \label{offset_plot}}
\end{figure}

The sky-plane position of 240P-B relative to 240P-A varies in a complicated way as a result of the changing viewing geometry (Figure \ref{offset_plot} and Table \ref{240P_data}). In order to estimate the time of separation and the separation speed, we measured the positions and fitted orbits to each component.  The positions were determined by centroiding within a 5$\times$5 pixel rectangular box and the program FindOrb (\url{https://www.projectpluto.com/}) was used to fit  Keplerian orbits. Table \ref{orbits} summarizes the results and compares solutions from NOT astrometry with solutions from the JPL Horizons Small Body Database Lookup (\url{https://ssd.jpl.nasa.gov/}). The NOT and JPL solutions for 240P-A are noticeably different for two reasons.  First,  the Horizons solution uses a much longer arc of observations extending over nine years vs.~only six months for the NOT data.  The long arc allows the inclusion of  non-gravitational acceleration parameters (not listed in Table \ref{orbits}) in the JPL orbit.  Second, the NOT data are relatively homogeneous in being obtained with a single telescope, camera, filter and measuring technique.  As a result, the root-mean-square uncertainty on the NOT orbit is three times smaller than on the JPL orbit for 240P-A.  

The JPL and NOT solutions for 240P-B are more consistent because the arc lengths  are  comparable (233 days vs.~168 days). Again, the NOT solution has a three times smaller root-mean-square error than the JPL orbit (0.20\arcsec~vs.~0.66\arcsec) presumably because of its greater observational consistency.    S. Nakano\footnote{\url{https://tinyurl.com/NEAT-B}} reports a solution for A1 and A2 in 240P-B, but finds that A1 is negative (i.e., a force directed towards the Sun), which would imply stronger sublimation from the night-side than from the day-side of the nucleus.  This solution seems unphysical, and we conclude that the non-gravitational acceleration of 240P-B is not well determined.   

 We used the Findorb solutions for both components in the NOT data to calculate the angular separation as a function of time, overplotted  in Figure \ref{offset_plot} as a solid red line. 
We also used the best-fit orbits to compute Sun-centered Cartesian coordinates for 240P-A and 240P-B, from which the linear separation was calculated as a function of time.  The resulting rate of increase of the separation in the period UT 2025 October 11 to UT 2026 March 28 (240P-B was not detected in the April 7 observation) is $\Delta V = 0.98\pm 0.02$ m s$^{-1}$.  

240P-B was first reported as an independent body on UT 2025 June 4 (DOY = 155; \cite{Jae25}).  However, the time of separation likely predates this first detection by years.  To see this we note that the linear distance between the two bodies was $\ell$ = 1.03$\times10^8$ m on UT 2026 January 1 (DOY 366).  Separation in 2025 June would imply separation speed $\sim$6 m s$^{-1}$, much larger than the measured value.  Instead, dividing the separation by the measured speed gives a time since separation  $t_{split} \sim  \ell/(d\ell/dt)$ = 1.05$\times10^8$ s (i.e., $\sim$3.4 year), corresponding to 2022 September, at which time the comet would have been near aphelion.  However, it is likely that $t_{split}$ is underestimated, because  differential non-gravitational acceleration likely increases the separation speed with time.   Indeed, it is tempting (although premature) to associate the splitting event with the long-lived brightening observed to start in the previous orbit (2018 November; \cite{Kel19}), at which time  $r_H \sim$ 2.5 au outbound.  Other examples of long-lived fragments exist in the short-period comet population (e.g., in comet 73P/Schwassmann-Wachmann 3, in which fragments survived from splitting in 1995 at least to 2001 (\cite{Boe02})). In any event, the split between 240P-A and 240P-B is not recent.

\section{Cause of Splitting}
\textbf{Tidal Forces:} 240P does not currently closely approach any planets or the Sun and therefore tidal splitting of its nucleus can be ruled out.   The comet did enter Jupiter's Hill sphere (radius 0.35 au) at its closest approach in 2008 but remained far outside the $\sim 3 R_J$ Roche radius interior to which tidal splitting of a fluid (strengthless) body is expected.  We confidently conclude that 240P is not tidally split.

\textbf{Impact:}   Impact with an asteroid is highly improbable for two reasons.  First, even in the ecliptic, disruptive collisions are rare on the ($\sim$0.5 Myr) median lifetime of Jupiter family comets  (the optical depth of the main belt is $\sim10^{-9}$). Second, 240P traverses asteroid belt distances while 0.75 au above the ecliptic, as a result of its highly inclined orbit. This makes collisions even less likely.  Moreover, impact disrupted asteroids produce swarms of fragments not  pairs, as shown by the existence of numerous collisionally produced dynamical families in the main belt. On these grounds, we discount the possibility that 240P is a collisionally split body.

\textbf{Subsurface Pressure Buildup:} Sublimation of  super-volatiles (e.g., carbon monoxide, carbon dioxide, conceivably nitrogen ice) or the sudden release of gas through the crystallization of buried amorphous ice \citep{Pri24} could, in principle, lead to sub-surface pressure build up causing an explosion and the ejection of a major fragment.  A problem with this hypothesis in the case of 240P is that the vertical scale of such an event is limited to a very thin surface skin as a result of the low thermal diffusivity, $\kappa$ [m$^2$ s$^{-1}$], of cometary material\footnote{\cite{Dav24} reports a thermal inertia $I$ = 30 MKS units, which, with $\rho_n$ = 500 kg m$^{-3}$ and $c_p$ = 10$^3$ J kg$^{-1}$ K$^{-1}$, corresponds to diffusivity $\kappa = 4\times10^{-9}$ m$^2$ s$^{-1}$.}.  The propagation of heat from the surface into the interior is limited to a few times the thermal skin depth, $d \sim (\kappa t)^{1/2}$, where $t$ is the duration of surface heating.  For example, if the diurnal timescale is $t$ = 1 day ($\sim10^5$ s), and with $\kappa = 4\times 10^{-9}$ m$^2$ s$^{-1}$, the thermal skin depth is $d \sim$ 2 cm.  Even on the orbital timescale, $t$ = 7.5 years ($\sim2\times10^8$ s), the skin depth reaches only $d \sim$ 1 m, which is far too small to excavate even the minimum 50 m radius 240P-B.  This vertical scale problem could be alleviated if the nucleus contains a deep network of interconnected voids through which pressure could build at depth, while maintaining a resistant surface crust \citep{Sam01}. However, in the absence of other evidence, such a configuration seems contrived. We conclude that subsurface pressure buildups might play a local, near-surface role in cometary outbursts, but they are less likely to split a large fragment from a nucleus. 

\textbf{Thermal Stresses:}
Thermal expansion at the surface can induce stresses sufficient to fracture cometary material, even at the $r_H \gtrsim$ 2 au heliocentric distances relevant to 240P.  These stresses have been suggested as a potential cause of cometary splitting events \citep{Tau87}.  The particulate nature and very low diffusivity of cometary material revealed by in-situ measurements reduce the likelihood that thermal fracture plays a role, however, and fracture by itself is incapable of launching bulk fragments against the gravity of the main nucleus.

\textbf{Rotational Instability:}  The surviving hypothesis is the rotational instability of the primary nucleus.  This explanation is attractive for two reasons.  First, it is already observationally established that the spins of sub-kilometer nuclei can quickly change in response to outgassing torques (\cite{Kok17}, \cite{Jew21}).  No new or unusual process need be invoked for the nucleus to be driven to rotational instability. Second, the low speed of separation between the 240P components ($\sim1$ m s$^{-1}$ now but probably much less originally) is slightly above but comparable to the escape speed from the primary ($\sim0.25$ m s$^{-1}$), as is naturally expected of rotational breakup.  Scaled from other comets, the spin-up timescale for a 400 m to 600 m radius nucleus with perihelion 1 $\le q \le$ 2 au is $\sim20 \le \tau \le$ 40 years (Jewitt 2021) and breakup can be expected within a small multiple of $\tau$ if the outgassing torque is sustained.

%While rotational instability is plausible in the context of other comets, the exact geometry of the breakup of 240P cannot be known.  The nucleus could have started as a single, weakly bonded rubble-pile aggregate and been driven to internal tension and breakup by outgassing.  Alternatively, it could have originally been a binary or bilobate (contact binary) body later separated by spin-up due to outgassing torques.  Several short-period comet nuclei have dumbell-like morphologies that likely indicate the contact between two pre-existing, more nearly spherical bodies which might be separated by outgassing torques. 

Sublimation mass loss itself  sets a separate  limit to the survival of 240P.  Represented as a sphere of density $\rho_n$ = 500 kg m$^{-3}$ and radius 500 m, 240P-A has mass  $M_n \sim 2.5\times10^{11}$ kg.  Given a mass loss per orbit $\Delta M_d \ge 1.5\times10^9$ kg (Section \ref{production}), the complete destruction of the nucleus is assured on a timescale $M_n/\Delta M_d \le$  160 orbits ($\lesssim$1200 years) assuming the current rate of activity to be sustained.  This exceeds the rotational instability timescale by at least an order of magnitude but is still very short ($\lesssim$ 1\%) compared to the $\sim$0.5 Myr dynamical lifetime  \citep{Lev94} of short period comets in the terrestrial planet region.  Short physical lifetimes underlie the dearth of known subkilometer comet nuclei, and undercut attempts to compare the small comet size distribution with the small Kuiper belt object population (where lifetimes against mass loss and spin-up are effectively infinite).  The short rotational destruction timescale of short-period comets also explains why the inner solar system holds only a modest population of potentially dormant comets (\cite{Kim14}, \cite{Bel15}) compared to the number expected if comets generally leave devolatilized remnants.

A direct observational test of the rotational instability hypothesis can be obtained by measuring the rotation period of 240P, which should be short.  The critical period, $P_c$ for a strengthless, spherical nucleus of density $\rho_n$ is $P_c = (3\pi/G\rho)^{1/2}$, where $G = 6.67\times10^{-11}$ N kg$^{-2}$ m$^2$ is the gravitational constant.  Substituting $\rho_n$ = 500 kg m$^{-3}$ gives $P_c$ = 4.7 hour, and a slightly larger $P_c$ would result if the nucleus is substantially aspherical.  No direct detection of the nucleus, or measurement of its rotational lightcurve, is possible with the existing near-perihelion data because of the dominance of the scattering cross-section by dust. Instead, the nucleus of 240P might be photometrically isolated when far from perihelion and less active, but then will be a faint and challenging target.

\clearpage
\section{SUMMARY}

We examined split comet 240P/NEAT finding a small nucleus which, like sub-kilometer nuclei in general, is subject to rapid destruction.  Specifically:

\begin{enumerate}

\item The  radius of 240P-A lies in the range $r_n$ = 400 m to 600 m. The radius of 240P-B is uncertain, with a possible value $r_n \sim$ 300 m and an absolute minimum $r_n$ = 50 m (density $\rho_n$ = 500 kg m$^{-3}$ assumed).

\item The separation ($\sim10^8$ m) and the rate of change of the separation ($\sim$1 m s$^{-1}$) of 240P-A from 240P-B point to splitting of the precursor nucleus at least three years before the present observations.   240P is a long-lived pair.

\item Dust from both components has effective mean particle size $\bar{a} \sim$ 50 $\mu$m, dust ejection velocity $V$ = 25 m s$^{-1}$ and production rates are peaked near perihelion at 130 kg s$^{-1}$ (240P-A) and 36 kg s$^{-1}$ (240P-B). The total dust mass loss from these two bodies, integrated over the $\sim$six month period of observation, is $\Delta M_d = 1.5\times10^9$ kg (240P-A) and $\Delta M_d = 2.3\times10^8$ kg (240P-B).

\item Splitting of the nucleus of 240P is unlikely to have occurred as a result of tides, impact, thermal fracture or subsurface pressure buildup.  Instead, rotational instability triggered by outgassing torques is the likely cause of splitting.  Both 240P-A and 240P-B are liable to further breakup and destruction by rotational instability on timescales that are shorter than the timescale due to sublimation mass loss.

%\item The absence of components smaller than B is consistent with the action of radiation pressure, given the long time since the splitting of the comet.

\end{enumerate}

\acknowledgments
We thank Anlaug Amanda Djupvik and the NOT observing team for making these observations possible, Bill Gray for providing access to his FindOrb orbit fitting software and the anonymous referee for comments on the manuscript. 
%% After the acknowledgments section, use the following syntax and the
%% \facility{} macro to list the keywords of facilities used in the research
%% for the paper.  Each keyword will be checked against the master list during
%% copy editing.  Individual instruments or configurations can be provided 
%% in parentheses, after the keyword, but they will not be verified.

%{\it Facilities:}  \facility{NOT, ZTF, ATLAS}.

%\clearpage

%% edition.

\clearpage

\clearpage
\begin{deluxetable}{lcccrrrrrcrrrr}
\tabletypesize{\scriptsize}
%\rotate
\tablecaption{Observation Log
\label{240P_geometry}}
\tablewidth{0pt}
\tablehead{\colhead{UT Date\tablenotemark{a}} & \colhead{Time\tablenotemark{b}} & \colhead{DOY$_{25}$\tablenotemark{c}} & \colhead{$r_H$\tablenotemark{d}}   & \colhead{$\Delta$\tablenotemark{e}} & \colhead{$\alpha$\tablenotemark{f}}  & \colhead{$\theta_{- \odot}$\tablenotemark{g}} & \colhead{$\theta_{-V}$\tablenotemark{h}} & \colhead{$\delta_{\oplus}$\tablenotemark{i}} & \colhead{$\nu$\tablenotemark{j}}     }

\startdata

%July 2 22:59 - 23:19& STANCAM & 4.450&  3.448&  2.5&  114.7&  95.0&   -0.9& 281.1  \\
Oct 11 & 04:28 - 04:52 & 284 & 2.192&  1.394&   19.8 &  279.1 & 234.1&   13.9&  333.8 \\
Oct 20 & 01:11 - 04:40 & 293 & 2.175&  1.315&   17.0&  284.9& 234.7&   13.0&  337.0 \\
Oct 30 & 02:13 - 02:23 & 303 & 2.159&  1.242&   13.3&  294.1& 235.3&   11.4&  340.7 \\
Nov 17 & 00:38 - 00:42 & 321 & 2.138 & 1.167&   6.9&  337.1& 235.6&   6.7&  347.4\\
Nov 25 & 01:29 - 01:32 & 329 & 2.131&  1.161&   6.7&   16.6& 235.3&    4.1&  350.5 \\
Dec 2 & 00:31 - 00:33 & 336 & 2.126&  1.169&    8.6&   42.9& 234.8&    1.7& 353.1 \\
Dec 18 & 00:23 - 00:26 & 352 & 2.122&  1.236&   15.2&   66.8& 232.9&   -3.5&  359.3 \\
Dec 28 & 01:44 - 01:47 & 362 & 2.123&  1.309&   19.1&   72.0& 231.6&  -6.3&  3.1& \\
Jan 11 & 00:10 - 00:14 & 376 & 2.129&  1.438&   23.1&   75.3& 230.1&   -9.1&    8.5\\
Jan 22  & 22:05 - 22:08 & 387 & 2.139&  1.567&   25.3&  76.6& 229.5&  -10.7&   13.0\\
Feb 5   & 22:09 - 22:12 & 401 & 2.155&  1.740&   26.7&   77.4& 229.7&  -11.5&   18.2 \\
Mar 28 & 20:24 - 20:27 & 452 & 2.259& 2.413&   24.4&   79.7& 238.0&   -8.8&   36.4 \\
Apr 7 & 20:25 - 20:30 & 462 & 2.286&  2.543&   23.1&   80.2& 240.7&   -7.6&   39.8 \\
\enddata

%% Text for table notes should follow after the \enddata but before
%% the \end{deluxetable}. Make sure there is at least one \tablenotemark
%% in the table for each \tablenotetext.

\tablenotetext{a}{UT Date in 2025/2026}
\tablenotetext{b}{Start times of the first and last observation, each night}
\tablenotetext{c}{Day of Year, 1 = UT 2025 January 1}
\tablenotetext{d}{Heliocentric distance, in au }
\tablenotetext{e}{Geocentric distance, in au }
\tablenotetext{f}{Phase angle, in degrees }
\tablenotetext{g}{Position angle of projected anti-solar direction, in degrees }
\tablenotetext{h}{Position angle of negative heliocentric velocity vector, in degrees}
\tablenotetext{i}{Angle of Earth from orbital plane, in degrees}
\tablenotetext{j}{True anomaly, in degrees}

\end{deluxetable}
%%%%%%%%%%%%%%%%%%%%%%%%%%%%%%%%%%%%%%%%%%%%%%%%%%%%%%%%%%%%%

\clearpage
\begin{deluxetable}{lcrrllc}
\tabletypesize{\scriptsize}
%\rotate
\tablecaption{Observations  of 240P
\label{240P_data}}
\tablewidth{0pt}
\tablehead{\colhead{Date\tablenotemark{a}} & \colhead{DOY$_{25}$\tablenotemark{b}} & \colhead{$\Delta RA$\tablenotemark{c}} & \colhead{$\Delta Dec$\tablenotemark{d}}   & \colhead{A/H$_A$/C$_A$/$\dot{M_d}$\tablenotemark{e}} & \colhead{B/H$_B$/C$_B$/$\dot{M_d}$\tablenotemark{f}}  & \colhead{$\Delta (B-A)$\tablenotemark{g}}       }

\startdata
%Jly 2 22:59 - 23:19& STANCAM & 4.450&  3.448&  2.5&  114.7&  95.0&   -0.9& 281.1  \\
%Oct 11 & 04:28 - 04:52 & 284 & 2.192&  1.394&   19.8 &  279.1 & 234.1&   13.9&  333.8 \\
Oct 11 & 284	&	-80.5	&	-56.1	&	14.30/11.08/1390/111	&	18.27/15.05/36/3	&	3.97$\pm$0.02\\
Oct 20 & 293	&	-86.9	&	-59.1	&	14.06/11.10/1368/109	&	17.72/14.75/47/4	&	3.66$\pm$0.02	\\
Oct 30 & 303	&	-93.1	&	-62.7	&	13.67/10.99/1503/120 &	17.01/14.33/69/6	&	3.34$\pm$0.02	\\
Nov 17 & 321	&	-100.2	&	-66.3	&	13.23/10.97/1534/123 &	15.28/13.02/232/18	&	2.05$\pm$0.02	\\
Nov 25 & 329	&	-99.9	&	-67.4	&	13.21/10.97/1529/122	&	14.84/12.60/341/27	&	1.63$\pm$0.02	\\
Dec 2 & 336	&	-97.8	&	-68.1	&	13.20/10.87/1675/134	&	14.64/12.31/445/36	&	1.44$\pm$0.02	\\
Dec 18 & 352	&	-88.8	&	-66.8	&	13.72/11.01/1474/118	&	15.40/12.69/314/25	&	1.68$\pm$0.02	\\
Dec 28 & 362	&	-81.7	&	-65.5	&	14.16/11.18/1267/101	&	16.14/13.16/205/16 &	1.98$\pm$0.02	\\
Jan 11 & 376	&	-72.1	&	-61.4	&	14.74/11.38/1050/84	&	16.35/12.99/238/19 &	1.61$\pm$0.02	\\
Jan 22 & 387	&	-65.3	&	-57.6	&	15.14/11.52/924/74	&	16.99/13.37/168/13	&	1.85$\pm$0.02	\\
Feb 5 & 401	&	-59.3	&	-52.0	&	15.62/11.70/782/62	&	17.76/13.84/109/9	&	2.14$\pm$0.02	\\
Mar 28 & 452	&	-51.1	&	-33.4	&	16.69/12.05/570/46	&	19.29/14.64/52/4	&	2.59$\pm$0.10	\\
Apr 7 & 462	&	-50.5	&	-29.7	&	17.00/12.27/466/37	&		--/	--/	--  &	 --\\
\enddata

%% Text for table notes should follow after the \enddata but before
%% the \end{deluxetable}. Make sure there is at least one \tablenotemark
%% in the table for each \tablenotetext.

\tablenotetext{a}{UT Date in 2025/2026 }
\tablenotetext{b}{Day of Year; 1 = UT 2025 January 1}
\tablenotetext{c}{Offset of B from A, arcseconds East}
\tablenotetext{d}{Offset of B from A, arcseconds North}
\tablenotetext{e}{Apparent magnitude/absolute magnitude/scattering cross-section/dust mass loss rate  (kg s$^{-1}$) of 240P-A within the 10$^4$ km aperture}
\tablenotetext{f}{Apparent magnitude/absolute magnitude/scattering cross-section (km$^2$)/dust mass loss rate (kg s$^{-1}$) of 240P-B within the 10$^4$ km aperture}
\tablenotetext{g}{Magnitude difference $\Delta(B-A)$ magnitudes}

\end{deluxetable}
%%%%%%%%%%%%%%%%%%%%%%%%%%%%%%%%%%%%%%%%%%%%%%%%%%%%%%%%%%%%%
\clearpage
\begin{deluxetable}{lllllc}
%\tabletypesize{\scriptsize}
%\rotate
\tablecaption{Osculating Orbital Elements\tablenotemark{a}
\label{orbits}}
\tablewidth{0pt}
\tablehead{\colhead{} & \colhead{240P-A} & \colhead{240P-A} & \colhead{240P-B}   & \colhead{240P-B}   \\
\colhead{Quantity\tablenotemark{b} } & JPL Horizons & ~This Work & JPL Horizons & ~This Work }

\startdata
%Jly 2 22:59 - 23:19& STANCAM & 4.450&  3.448&  2.5&  114.7&  95.0&   -0.9& 281.1  \\
%Oct 11 & 04:28 - 04:52 & 284 & 2.192&  1.394&   19.8 &  279.1 & 234.1&   13.9&  333.8 \\
a      & 3.8615010(2) & 3.8626(7) &  3.86088(3) & 3.8624(7)           \\
e      & 0.4503324(1) & 0.45074(9) &  0.450469(4) & 0.4507(7)           \\
%q      & 2.1225420(6) & 2.12158(7) &  2.121675(6) & 2.12165(8)           \\
i      &23.53697822(8) & 23.533(1) &  23.53587(8) & 23.533(2)           \\
$\Omega$   & 74.91442(1) & 74.9099(8) & 74.91181(4) & 74.9101(8)           \\
$\omega$   & 352.02572(4) & 352.07(1) & 352.072(5) & 352.07(1)           \\
M      &291.04666(2) & 14.028(6)  & 358.313(1) & 12.726(6)           \\
$T_P$  & Dec-19.8666(1) & Dec 19.94(3) & Dec-19.983(1)  & Dec 19.98(3)           \\
%Source & JPL Horizons & This Work & JPL Horizons & This work \\
Arc   & 3218 & 178 & 233 & 168 \\
Residual & 0.66  &  0.09  & 0.66 & 0.20 \\

Reference      & (JPL K252/41)  & --- & (JPL 17)  &    ---        \\
%N$_{obs}$      &   4039             &  13   & 1454   &   12   \\

%e      &              &           &              &            \\

\enddata

%% Text for table notes should follow after the \enddata but before
%% the \end{deluxetable}. Make sure there is at least one \tablenotemark
%% in the table for each \tablenotetext.
\tablenotetext{a}{For brevity the numbers in parentheses show  1$\sigma$ uncertainties in the last digit (e.g.,  3.14159(6) = 3.14159$\pm$6)}
\tablenotetext{b}{ $a$ = semimajor axis (au),  $e$ = eccentricity, $i$ = inclination (degree), $\Omega$ = longitude of ascending node (degree), $\omega$ = argument of perihelion (degree), $M$ = mean anomaly (degree), $T_P$ = date of perihelion (2025), Arc = Arc length (days), Root mean square residual of fitted orbit (arcsecond)}

\end{deluxetable}
%%%%%%%%%%%%%%%%%%%%%%%%%%%%%%%%%%%%%%%%%%%%%%%%%%%%%%%%%%%%%
\clearpage
\begin{deluxetable}{lllllc}
%\tabletypesize{\scriptsize}
%\rotate
\tablecaption{Summary Table
\label{summary}}
\tablewidth{0pt}
\tablehead{\colhead{Quantity} & \colhead{240P-A} & \colhead{240P-B}  }

\startdata
%Jly 2 22:59 - 23:19& STANCAM & 4.450&  3.448&  2.5&  114.7&  95.0&   -0.9& 281.1  \\
%Oct 11 & 04:28 - 04:52 & 284 & 2.192&  1.394&   19.8 &  279.1 & 234.1&   13.9&  333.8 \\
Radius\tablenotemark{a} [m]                  & 500$\pm$100 &  $\ge$ 50\\
Mass\tablenotemark{b}   [kg]                 & $2.6^{+1.9}_{-1.3}\times10^{11}$ & $\ge2.3\times10^8$ \\
$\dot{M}$\tablenotemark{c} [kg s$^{-1}$]     & 130 &  36 \\
$\Delta M_d$\tablenotemark{d} [kg orbit$^{-1}$]       & $1.5\times10^9$  & $2.3\times10^8$ \\
$\tau = M t /\Delta M_d$\tablenotemark{e} [year]   & 1200 & $\ge$7 \\
$\tau_s $\tablenotemark{f} [year]            & 25 &  $<$ 1\\

\enddata

%% Text for table notes should follow after the \enddata but before
%% the \end{deluxetable}. Make sure there is at least one \tablenotemark
%% in the table for each \tablenotetext.
\tablenotetext{a}{Radius of equal mass sphere}
\tablenotetext{b}{Nucleus mass assuming density $\rho_n$ = 500 kg m$^{-3}$}
\tablenotetext{c}{Peak dust mass loss rate}
\tablenotetext{d}{Mass loss per orbit}
\tablenotetext{e}{Mass loss lifetime, $t$ = orbital period in years}
\tablenotetext{f}{Spin-up timescale}

\end{deluxetable}
%%%%%%%%%%%%%%%%%%%%%%%%%%%%%%%%%%%%%%%%%%%%%%%%%%%%%%%%%%%%%

\clearpage

%%%%%%%%%%%%%%%%%%%%%%%%%%%%%%%%%%%%%%%%%%%%%%%%%%%%%%%%%%%%%%%%%%%
%%%%%%%%%%%%%%%%%%%%%%%%%%%%%%%%%%%%%%%%%
%%%%%%%%%%%%%%%%%%%%%%%%%%%%%%%%%%%%%%%%%
%%%%%%%%%%%%%%%%%%%%%%%%%%%%%%%%%%%%%%%%%
\clearpage

\clearpage

%\begin{figure}

\clearpage

\clearpage

\clearpage

\clearpage

\clearpage
%%%%%%%%%%%%%%%%%%%%%%%%%%%%%%%%%%%%%%%%%
%%%%%%%%%%%%%%%%%%%%%%%%%%%%%%%%%%%%%%%%%
%%%%%%%%%%%%%%%%%%%%%%%%%%%%%%%%%%%%%%%%%
\clearpage

%%%%%%%%%%%%%%%%%%%%%%%%%%%%%%%%%%%%%%%%%
\clearpage

%%%%%%%%%%%%%%%%%%%%%%%%%%%%%%%%%%%%%%%%%
%%%%%%%%%%%%%%%%%%%%%%%%%%%%%%%%%%%%%%%%%
%%%%%%%%%%%%%%%%%%%%%%%%%%%%%%%%%%%%%%%%%

\end{document}